\documentstyle[preprint,prl,aps]{revtex}
\begin{document}
\draft

\preprint{\vbox{\hbox{May 1996}\hbox{rev. Aug. 1996}\hbox{IFP-728-UNC}}}


\title{A New Approach to the Family Structure.}
\author{\bf  Otto C. W. Kong}
\address{Institute of Field Physics, Department of Physics and Astronomy,\\
University of North Carolina, Chapel Hill, NC  27599-3255}
\maketitle

\begin{abstract}
In this letter, we introduce a new approach to formulate the 
family structure of the  standard model. Trying to mimic the highly 
contrained representation structure of the standard model while
extending the symmetry, we propose a 
$SU(4)\otimes SU(3)\otimes SU(2)\otimes U(1)$ symmetry with a
SM-like chiral spectra basically "derived" from the gauge anomaly constraints.
Embedding the SM leads to 
$SU(4)_A\otimes SU(3)_C\otimes SU(2)_L\otimes U(1)_X$ models, which upon the 
$SU(4)_A\otimes U(1)_X \longrightarrow U(1)_Y$ symmetry breaking, 
gives the three families  naturally as a result. A specific model obtained from
the approach is illustrated.  The model, or others from our approach, holds promise of a very 
interesting phenomenology. We sketch some of the results here. 
An interesting possiblity of supersymmetrizing the model with the EW-Higgses
already in the spectrum is noted. A comparison with  other approaches is also  discussed.

\end{abstract} 


\newpage 

In this letter, we introduce a new approach to the much pursued, 
however still mysterious, family problem. Trying to mimic the highly 
contrained group representation structure of the standard model while
extending the symmetry, we describe a specific
$SU(4)_A\otimes SU(3)_C\otimes SU(2)_L\otimes U(1)_X$ 
model which, upon symmetry
breaking, gives the three-family standard model together with some 
extra vector-like states. The fully chiral fermion spectrum of the model,
is largely "derivable" from the gauge anomaly constraints, like the SM.
Details of the representation content of the SM-like chiral spectrum
are given in table 1. 
To set the stage and facilitate comparison,
we first note the major available approaches, before we
describe our new alternative.

The matter content of the phenomenologically very successful 
standard model(SM) consists of three sets (families) of 15 chiral fermion
states of identical quantum numbers. In addition, a scalar doublet
Higgs is needed to break the electroweak(EW) symmetry and give the fermions
masses. The existence of the scalar leads 
to the hierarchy problem, which is widely believed to be addressed by
supersymmetry(SUSY). 
The representation structure
of a single family of chiral fermions  is very strongly constrained by
the requirement of cancellation of all  gauge anomalies, making it
easily "derivable" once some simple assumptions are taken\cite{sm1}.
However, the existence of {\it three} families, with the great hierarchy of
masses among the fermions after EW-symmetry breaking, remains a
mystery.

There have been many attempts on explaining the family structure since 
the late 70's. The simplest approach is to introduce an extra family
or horizontal symmetry commuting with the standard model
group or its (vertical) unification group\cite{ghs}. It is desirable
to have a gauged horizontal symmetry since anomaly constraints reduce
much of the arbitrariness in the model building exercises. Moreover, 
analysis of gravitational effect\cite{gg} 
casts a strong doubt on the consistence
of global horizontal symmetry models, at least for those that have a
high symmetry breaking scale. Recently, there are a lot of activities
on this kind of model building\cite{q2n}, aiming at obtaining 
phenomenologically viable textures for the quark, and the less constrained
squark, mass matrices\cite{tex}.

There are more non-trivial ways of putting the family structure into
a gauge symmetry. For instance the very nice embedding of the one
family SM  into a $SU(5)$ unification\cite{su5} had motivated
further unification models that lead to three chiral families\cite{79,s9,s7}.
An illustrative example\cite{s9} has the following structure: 
the symmetry breaking is assumed to be 
$SU(9) \longrightarrow SU(5) \longrightarrow SU(3)_C\otimes SU(2)_L\otimes U(1)_Y$ ;
an anomaly free fermion content,  $9{\bf [9,8]} + {\bf [9,3]}$, is taken
which then results in  three chiral $SU(5)$ families. 
There is no apparant way to build the fermion mass hierarchy
among the families. Moreover, the family structure is built in above
the remote unification scale. Unlike the horizontal symmetry approach,
this one is not much pursued for quite some time.

Yet another interesting alternative is proposed by Frampton\cite{331}
relatively recently, called the 331-model. There
$SU(3)_C\otimes SU(2)_L\otimes U(1)_Y$ of the SM is embedded into
$SU(3)_C\otimes SU(3)_L\otimes U(1)_X$ with a nontrivial hypercharge
embedding. The three families are put in separately; however, 
the third family has  a representation structure different
from that of the first two and gauge anomalies cancelled among the three families. Extra structure is needed, to 
produce the symmetry breaking and give the exotic states reasonably
large mass. 

Now, let us take a look at the elegance of the SM representation
structure for a single family. For example: 
\begin{itemize}
\item
We can start by introducing
the simplest multiplet that transforms nontrivially under each of 
the component group factors,
namely a ${\bf (3,2,1)}$. (Here fixing the hypercharge as 1 
just corresponds to a arbitrary choice of normalization.) 
\item
To cancel the
$SU(3)$ anomaly, two ${\bf \bar{3}}$'s are needed. Keeping with the chiral structure, 
we have to use a ${\bf (\bar{3},1,x)}$ and a ${\bf (\bar{3},1,y)}$.
\item
Next, cancellation of the global-$SU(2)$ anomaly dictates the inclusion of
an extra doublet,  ${\bf (1,2,a)}$. 
\item
We still have to cancel all the $U(1)$-anomalies.
We have $a=-3$ from the $[SU(2)]^2U(1)$ anomaly constraint.
$x$ and $y$ then has to satisfy  the three remaining contraints and no solution can be
found. If we then allow one singlet state, ${\bf (1,1,k)}$, we have three equations
for three unknowns giving a unique solution, the SM hypercharge 
assignment $(x, y = 2, -4; k=6)$. Notice that the solution {\it a priori} may not give a
set of rational numbers. The triviality of solution here is a bit
deceiving.
\end{itemize}

We want to try to mimic the above feature in a extended symmetry that
can incorporate the three families naturally. Enlarging one of the 
component group factors 
apparently does not work. We consider adding one more factor. Then 
$SU(4)\otimes SU(3)\otimes SU(2)\otimes U(1)$ suggests 
itself as the most natural candidate.
\begin{itemize}
\item
We start with a ${\bf (4,3,2,1)}$. Develop along the same lines as above, 
we have  the list:
\begin{center} 
${\bf (4,3,2,1)}, {\bf (\bar{4},\bar{3},1,x)}, {\bf (\bar{4},1,2,y)}, 
{\bf (\bar{4},1,1,z)},$  \\
${\bf (1,\bar{3},2,a)}, {\bf (1,\bar{3},1,b)}, {\bf (1,\bar{3},1,c)},$ \\
${\bf (1,1,2,k)}, {\bf (1,1,1,s)}$. 
\end{center}
Notice that all gauge anomalies not involving the $U(1)$ group cancel;
in particular, there is an equal number of ${\bf 4}$'s and 
${\bf \bar{4}}$'s, or  ${\bf 3}$'s and ${\bf \bar{3}}$'s, 
and a even number of  ${\bf 2}$'s.
\item
If we asume we have a 
$SU(4)_A\otimes SU(3)_C\otimes SU(2)_L\otimes U(1)_X$ , the list
has basically what it takes to give the three chiral families after
$SU(4)_A\otimes U(1)_X$ breaks into $U(1)_Y$. Replacing the $SU(4)$ by 
other $SU(N)$ groups may also be considered.  $N<4$,
cannot  accomodate  three families, nor can $N>6$. 
$N=5$ and $N=6$ present viable alternatives 
when given a bit "un-natural" representation content (see table 2).
One can also consider a nontrivial embedding of $SU(3)_C$ or $SU(2)_L$.
That would not work either, at least not for $N=4, 5$ and $6$\cite{532}.
\item
Finding a set of $X$-charge assignments that can cancel all
the related anomalies is highly nontrivial. While the author does have
a successful solution, the approach may not be very fruitful.
We have another issue on the agenda: building the correct hypercharge
embedding that enables us to identify the SM families.
\item
There are three independent $U(1)$ factors in $SU(4)$. To embed 
$U(1)_Y$, we can  identify it as a linear combination of the
three and the extra  $U(1)_X$. Remember that we want a three-family SM. For instance, there are four 
quark doublets contained in ${\bf (4,3,2,1)}$. We want the correct 
hypercharge for three of them; and  the fourth better form a vector-like 
pair with ${\bf (1,\bar{3},2,a)}$.
Assuming SUSY, the chiral multiplets contain also scalar states. We
can then assume  a hypercharge invariant VEV for 
a scalar state in ${\bf (\bar{4},1,1,z)}$, 
that breaks $SU(4)_A\otimes U(1)_X$ into for example 
$SU(3)_H\otimes U(1)_Y$ with the $SU(3)_H$ as a horizontal symmetry
for the SM. This fixes the quark doublets. Putting in all the 
requirements from  the SM $U(1)_Y$ assignments,
there is a unique solution for the set of $X$-charges and a
specific $U(1)_Y$ definition as a linear combination of 
$U(1)_X$ and the other $U(1)$ from $SU(4)_A$, up to a two fold
ambiguity in embedding the quark singlets.
\item
Finally, we have to check all the $U(1)_X$-anomalies. It sounds like 
we need a miracle to have  all the gauge anomalies just cancelled. It does  {\it not}
work! However, a small modification of the rules of the game does give
us what we want: a consistent gauge model of three SM families
together with the extra  states all forming vector-like pairs after the
symmetry breaking! 
We just need to add a ${\bf (6,1,1,p)}$, which is free of 
$SU(4)$-anomaly, and some more singlets. 
The detailed structure of a specific model is listed in table 1.
\end{itemize}

We just sketched above our model construction approach. 
Apart from the two embedding alternatives,
there is still some flexibility in modifying the leptonic part of the spectrum.
These, together with a detailed discussion
of our construction and some phenomenological features, we leave to another publication\cite{729}.  We give below an outline of some of   the results presented there.
The specific model is used to illustrate the basic features of models from
our approach.

We have three SM quark doublets from ${\bf (4,3,2,1)}$, which also 
contains, together with ${\bf (1,\bar{3},2,-10)}$, a fourth vector-like
quark doublet($Q^{'}$) with electric charges $(-1/3, -4/3)$.
The singlet quark  states have interesting embeddings. The three
$\bar{u}$'s and one $\bar{d}$ are in ${\bf (\bar{4},\bar{3},1,5)}$
while the other two $\bar{d}$'s are the ${\bf (1,\bar{3},2,-4)}$'s. This
difference in structure between the up- and down-sector might explain 
the expected difference in the textures of the two mass matrices\cite{tex}
and hence the very existence of the CKM-matrix.

The scalar $\phi_0 = {\bf (\bar{4},1,1,9)}$ with the natural
$SU(4)_A\otimes U(1)_X$ breaking VEV
\begin{equation}
$$\left \langle \phi_0 \right \rangle =\left(\begin{array}{cccc}
0 & 0 & 0 & v
\end{array}\right) $$ 
\end{equation}
can actually be used to define the remnant $SU(3)_H$ and $U(1)_Y$ symmetry.
A possible scheme of quark mass generation then involves taking the Higgs doublets
from a  $\Phi = {\bf (15,1,2,-6)}$ together with extra SM singlet scalars from  
two ${\bf (\bar{4},1,1,-3)}$'s, denoted by $\phi_a$ ($a=1$ or $2$), with natural VEVs
\begin{equation}
$$\left \langle \phi_1 \right \rangle =\left(\begin{array}{cccc}
v_1 & 0 & 0 & 0
\end{array}\right) $$ ,
\hspace{.5in}
$$\left \langle \phi_2 \right \rangle =\left(\begin{array}{cccc}
v_1^{'} & v_2 & 0 & 0
\end{array}\right) $$ .
\end{equation}
A mass term for   $\Phi$   of the form\cite{SH}
\[
C_{ab} \phi_{ai} \phi_b^{\dag  j} \Phi^k_j \Phi^{\dag i}_k
\]
then gives masses to all component doublets of the scalar except three, two of which 
contain zero electric charge states that can function as the EW-breaking Higgses.
The two Higgses each couples directly  to only one  of SM quark; they give
mass to only the top and the bottom. Hence,
FCNC constraints\cite{fcsc} could be satisfied by assuming 
$v_i \geq 200TeV$. FCNC's from the extra, nine, neutral gauge bosons would 
also  be under control then\cite{FCNC}.  Smaller masses for the lighter quarks
could be generated through other secondary mechanism or an enriched scalar
sector, giving naturally hierarchial quark mass matrices. For example, in a non-SUSY
setting,  the given set of scalar VEVs can be combined to give suppressed effective
mass terms to the second family, leaving the first family masses to be generated by
radiative effects.

The third light doublet consists of Higgses
of electric charges 1 and 2. The doubly charged Higgs is a novel prediction of 
the model. They couple the SM quarks to the $Q^{'}$. The Higgs that is responsible 
for the top mass also gives mixing between the fourth down-type quark in  $Q^{'}$
and the bottom. 
This mixing fixes the $R_b$-anomaly\cite{rbc}
if $M_{Q^{'}}\sim 715GeV$. The scenario   may be possible if 
value of $M_{Q^{'}}$ is suppressed by small (effective) Yukawa coupling. 

Upon QCD confinement, the new quarks will lead to
interesting new mesonic and baryonic states. Notice that none of these
has a fractional charge.

For the leptonic doublets, we have a $\bar{L}$ from ${\bf (\bar{4},1,2,3)}$
and four $L$'s from the rest of  ${\bf (\bar{4},1,2,3)}$
and the ${\bf (1,1,2,6)}$. In a SUSY scenario, the extra pair of vector-like doublets
might be identified as the Higgs(ino) doublets, instead of using $\Phi$ above,  hence making our SM-like chiral spectrum self-contained; as the  $\phi_0$ scalar is already
in the spectrum, as remarked above. The model then has lepton-number violation
in a way similar to some recent analysis\cite{lnv}.
The details of the quark and lepton mass generations, and the scalar masses including the soft SUSY-breaking parts have however to be analyzed within the framework of
our extended symmetry to see if the model could then be made
consistent and realistic.

The SM-singlet sector is much richer. There is a
right-handed neutrino state($N$) in  the ${\bf (\bar{4},1,1,9)}$, which can
develop Majoarna mass invariant under the SM symmetry. 
The right-handed neutrino may give rise to  neutrino masses
through the see-saw mechanism. That may retrospectively be 
used to fix the symmetry breaking scale. There are also three extra
vector-like singlet states of both electric charge 1($E/\bar{E}$) and 2($S/\bar{S}$). 
The former will be  involved in mixing with the charged-leptons in the corresponding mass 
matrix.

While there is no
obvious gauge group unification, string-type gauge coupling unification
may not be ruled out. With SUSY assumed and the SM-like chiral
spectrum taken without extra supermultiplets, the coefficients for the first order $\beta$-functions are given by 
$ (b_4,b_3,b_2,b_1)=\frac{1}{16\pi^2}(-5,-1,4,233/24)$
 where we have normalized the
$X$-charge by 1/24. We can see that the model just maintains 
the $SU(3)_C$ asymptotic freedom, a feature shared also by the $SU(4)_A$
component.  We have to break $SU(4)_A$ before it confines to get the 
correct number of families. Assuming gauge coupling unification, this could
be used to set a limit for the scale of the symmetry breaking.  However, extra supermultiplets incorporating the symmetry breaking scalars, such as $\Phi$
used above, would have to be included in a realistic model.  They  would have very 
strong effect on the all the coefficient except $b_3$, easily removing the 
asymptotic freedom of  $SU(4)_A$ for instance. The analysis is very important
but cannot be done without a fully realistic model. this we will leave for further investigations.

Finally, we compare our approach with other formulations of a 
three-family SM. From the descriptions above, the major features are
obvious. Our construction approach is unique. It has a  SM-like chiral spectrum which is
largely "derived", following similar features of the one family SM, and {\it yield
three families as a natural result}.
The model obtained from the approach,
however, looks  in some sense a hybrid of the three
different major approaches used by previously authors. It has part of the extra symmetry
similar to the horizontal symmetries. The major difference between the latter
and our symmetry structure as an extension of the SM symmetry is that the hypercharge embedding is  nontrivial in our case, leading to existence of extra charged gauge bosons. Like the other two approaches also mentioned at the beginning,
we start with a fully chiral  high energy fermion
content, which then gives rise to the correct low energy chiral
fermion content in a nontrivial way. However, it is not a
grand unification type model. There is a possibility of embedding
the  $SU(4)_A\otimes SU(3)_C\otimes SU(2)_L\otimes U(1)_X$ group into a
$SU(9)$ or some other simple groups of higher rank. No such unification
scheme is apparent for incorporating the chiral spectrum.  Because it is not 
tied up with unification, the
symmetry breaking scale may not be remote, giving it plausible
interesting phenomenology may be accessible by future experimental
machines, like the 331-model\cite{331}. The 331-model also shares 
a nontrivial hypercharge embedding.
Our  model contains 84 chiral fermionic states. This is to be compared with
the 165 states from the $SU(9)$ model sketched above.  
Another similar minimal three family model with a $SU(7)$ symmetry has
133 states\cite{s7}. With only 63 states, the 331-model is still a bit more economic.
However, the 84-state  SM-like chiral spectrum, shares, to a great extent, the
elegant structure of that of one SM family. 
Further investigations of the properties  of the model, 
or others of the type, worth serious efforts.

{\it Acknowledgement:} The author would like to thank P.H. Frampton for reading the
first version of the manuscript and for valuable comments and suggestions.

This work was supported in part by the U.S. Department of 
Energy under Grant DE-FG05-85ER-40219, Task B.\\


\bigskip
\bigskip

\newpage

{\bf Table Caption.}\\

\bigskip
\bigskip

Table 1: The $SU(4)_A\otimes SU(3)_C\otimes SU(2)_L\otimes U(1)_X$
model -- representation structures, anomaly cancellations and
the SM embedding. \\
The hypercharge (differ by a factor of 6 from the standard 
normalization)  is given   by  $Y=3/2 Z -1/2 X$ where
$Y, Z$ and $X$ are charges of correspondent $U(1)$ groups, with
$U(1)_Z\otimes SU(3)_H \subset SU(4)_A$. $Q, \bar{u}, \bar{d}$ denote the 
quark multiplets; $L$ and $E$ multiplets with the quantum numbers of
leptonic doublets and singlets; $S, \bar{S}$,  $Q^{'}$ and
$\bar{Q^{'}}$, the vector-like singlets and quark doublets; and 
$N$ a right-handed neutrino state.

\bigskip

Table 2: Suggestive representation structures from
the standard model to 
$SU(N)_A\otimes SU(3)_C\otimes SU(2)_L\otimes U(1)_X$,
$N=4, 5$ and 6, with three families. \\
Here we suppress the $U(1)_X$-charges.

\newpage


Table 1: The $SU(4)_A\otimes SU(3)_C\otimes SU(2)_L\otimes U(1)_X$
model -- representation structures, anomaly cancellations and
the hypercharge embedding.

\vspace{.3in}

\scriptsize

\begin{tabular}{|c|c|r|r|r|r|r|cc|}
\hline\hline
$SU(4)_A\otimes SU(3)_C\otimes SU(2)_L$ Rep. & $U(1)_X$ &
\multicolumn{5}{|c|}{Gauge anomalies} &  \multicolumn{2}{|c|}{$U(1)_Y$ states } \\ 
\hline 
&  &	$U(1)$-grav. & $[SU(4)]^2U(1)$	& $[SU(3)]^2U(1)$ &  $[SU(2)]^2U(1)$ &	$[U(1)]^3$ &				& \\
\hline								  		
${\bf (4,3,2)}$		&	{\bf 1}  &	24  &	6  &	8 &	12 &	24	& 	3\ {\bf 1}($Q$)	   	& {\bf -5}($Q^{'}$) \\
${\bf (\bar{4},\bar{3},1)}$ &	{\bf 5}  &	60  &	15 & 	20 &	 &	1500	& 	3\ {\bf -4}($\bar{u}$)	& {\bf 2}($\bar{d}$) \\
${\bf (\bar{4},1,2)}$	&	{\bf 3}  &	24  &	6  &	&	12 &	216	& 	3\ {\bf -3}($L$)	& {\bf 3}($\bar{L}$) \\
${\bf (\bar{4},1,1)}$	&	{\bf 9}  &	36  &	9  &	&	 &	2916	& 	3\ {\bf -6}($\bar{E}$)	& {\bf 0}($N$) \\
${\bf (6,1,1)}$		&	{\bf -18} &	-108 &	-36 &	&	 &	-34992	& 	3\ {\bf 6}($E$)	& 3\ {\bf 12}($S$) \\ \hline
${\bf (1,\bar{3},2)}$	&	{\bf -10} &	-60 &	&	-20 &	-30 &	-6000	& 	\multicolumn{2}{|c|}{{\bf 5}($\bar{Q}^{'}$)} \\
${\bf (1,\bar{3},1)}$	&	{\bf -4} & 	-12 &	&	-4  &	 &	-192	& 	\multicolumn{2}{|c|}{{\bf 2}($\bar{d}$)} \\
${\bf (1,\bar{3},1)}$	&	{\bf -4} &	-12 &	&	-4  &	 &	-192	& 	\multicolumn{2}{|c|}{{\bf 2}($\bar{d}$)} \\ \hline
${\bf (1,1,2)}$		&	{\bf 6}  &	12  &	&	&	6  &	432	&	\multicolumn{2}{|c|}{{\bf -3}($L$)} \\
$3\ {\bf (1,1,1)}$ 	&	{\bf 24} &	72  &	&	&	&	41472	&	\multicolumn{2}{|c|}{3\ {\bf -12}($\bar{S}$)} \\ 
$3\ {\bf (1,1,1)}$ 	&	{\bf -12} &	-36 &	&	&	&	-5184	&	\multicolumn{2}{|c|}{3\ {\bf 6}($E$)} \\	\hline
\multicolumn{2}{|r|}{\it Total}    	&	0   &	0  &	0  &	0  & 	0	& 				& \\		
\hline\hline
\end{tabular}

\normalsize

\vspace{1.5in}

\newpage

Table 2: Suggestive representation structures from
the standard model to 
$SU(N)_A\otimes SU(3)_C\otimes SU(2)_L\otimes U(1)_X$,
$N=4, 5$ and 6, with three families.

\vspace{.3in}

\begin{tabular}{|c|c|c|c|}
\hline\hline
 SM				& $N=4$				& $N=5$				& $N=6$		\\
\hline\hline
				& ${\bf (4,3,2)}$		& ${\bf (5,3,2)}$		& ${\bf (6,3,2)}$		\\	
				& ${\bf (\bar{4},\bar{3},1)}$	& ${\bf (\bar{5},\bar{3},1)}$	& ${\bf (\bar{6},\bar{3},1)}$	\\
				& ${\bf (\bar{4},1,2)}$		& ${\bf (\bar{5},1,2)}$		& ${\bf (\bar{6},1,2)}$		\\
				& ${\bf (\bar{4},1,1)}$		& ${\bf (\bar{5},1,1)}$		& ${\bf (\bar{6},1,1)}$		\\
\hline
${\bf (3,2)}$			& ${\bf (1,\bar{3},2)}$		& ${\bf (1,\bar{3},2)}$		& ${\bf (1,\bar{3},2)}$		\\
				&				& ${\bf (1,\bar{3},2)}$		& ${\bf (1,\bar{3},2)}$		\\
				&				&				& ${\bf (1,\bar{3},2)}$		\\
${\bf (\bar{3},1)}$		& ${\bf (1,\bar{3},1)}$		& ${\bf (1,\bar{3},1)}$ 	& 				\\
${\bf (\bar{3},1)}$		& ${\bf (1,\bar{3},1)}$		&				&				\\
\hline
${\bf (1,2)}$			& ${\bf (1,1,2)}$		&				& ${\bf (1,1,2)}$		\\
${\bf (1,1)}$			& ${\bf (1,1,1)}$		& ${\bf (1,1,1)}$		& ${\bf (1,1,1)}$		\\	
\hline\hline
$\# Q=1(\times 3)$	& $\# Q=4-1$ & $\# Q=5-2$ & $\# Q=6-3$ \\
$\# \bar{q}=2(\times 3)$ & $\# \bar{q}=4+2$ & $\# \bar{q}=5+1$ & $\# \bar{q}=6$ \\		 
$\# L=1(\times 3)$	& $\# L=4\pm 1$ & $\# L=5$ & $\# L=6\pm 1$ \\
\hline\hline
\end{tabular}

\end{document}